\documentclass[%
preprint, aps, prx,
groupedaddress,
superscriptaddress,
longbibliography,
 amsmath,amssymb,
 showpacs
floatfix
]{revtex4-2}
\usepackage{graphicx}
\usepackage{makecell,booktabs}
\usepackage{dcolumn}
\usepackage{bm}
\usepackage{hyperref}
\hypersetup{
    colorlinks=true,
    urlcolor= blue,
    citecolor=blue,
linkcolor= blue}
\usepackage{xcolor}

\usepackage{xr}

\makeatletter
\newcommand*{\addFileDependency}[1]{
  \typeout{(#1)}
  \@addtofilelist{#1}
  \IfFileExists{#1}{}{\typeout{No file #1.}}
}
\makeatother

\newcommand*{\myexternaldocument}[1]{%
    \externaldocument{#1}%
    \addFileDependency{#1.tex}%
    \addFileDependency{#1.aux}%
}
\myexternaldocument{Islam_supplementary}

\renewcommand{\vec}[1]{{\textbf{\textit{#1}}}}
\newcommand{\figref}[1]{Fig.\,\ref{#1}}

\newcommand{\CdAs}{$\mathrm{Cd_{3}As_{2}}$}
\newcommand{\InMnAs}{$\mathrm{In_{1-x}Mn_{x}As}$}

\usepackage[skip=5pt, indent=10pt]{parskip}
\begin{document}

\title{Rashba spin splitting-induced topological Hall effect in a Dirac semimetal-ferromagnetic semiconductor heterostructure}

\author{Saurav Islam}
\email{ski5160@psu.edu}
 \affiliation{Department of Physics, Pennsylvania State University, University Park, Pennsylvania 16802, USA}
\author{Emma Steinebronn}
 \affiliation{Department of Physics, Pennsylvania State University, University Park, Pennsylvania 16802, USA}
 \author{Kaijie Yang}
 \affiliation{Department of Physics, Pennsylvania State University, University Park, Pennsylvania 16802, USA}
\author{Bimal Neupane}
\affiliation{Department of Physics, University of North Texas, Denton, Texas 76203}
\author{Juan Chamorro}
\affiliation{Department of Chemistry, Johns Hopkins University, Baltimore, Maryland 21218, USA}
 \author{Supriya Ghosh}
\affiliation{%
Department of Chemical Engineering and Materials Science, University of Minnesota, Minneapolis, Minnesota 55455, USA
}%
\author{K. Andre Mkhoyan}
\affiliation{%
Department of Chemical Engineering and Materials Science, University of Minnesota, Minneapolis, Minnesota 55455, USA
}%
\author{Tyrel M. McQueen}
\affiliation{Department of Chemistry, Johns Hopkins University, Baltimore, Maryland 21218, USA}
\author{Yuanxi Wang}
\affiliation{Department of Physics, University of North Texas, Denton, Texas 76203}
\author{Chaoxing Liu}
\email{cxl56@psu.edu}
\affiliation{Department of Physics, Pennsylvania State University, University Park, Pennsylvania 16802, USA}
\author{Nitin Samarth}
\email{nxs16@psu.edu}
\affiliation{Department of Physics, Pennsylvania State University, University Park, Pennsylvania 16802, USA}
\affiliation{Department of Materials Science \& Engineering, Pennsylvania State University, University Park, Pennsylvania 16802, USA}

\begin{abstract}

We use a concerted theory-experiment effort to investigate the formation of chiral real space spin texture when the archetypal Dirac semimetal Cd$_3$As$_2$ is interfaced with \InMnAs, a ferromagnetic semiconductor with perpendicular magnetic anisotropy. Our calculations reveal a nonzero off-diagonal spin susceptibility in the \CdAs~layer due to the Rashba spin-orbit coupling from broken inversion symmetry. This implies the presence of a Dzyaloshinskii-Moriya interaction between local moments in the \InMnAs~layer, mediated by Dirac electrons in the vicinal \CdAs~layer, potentially creating the conditions for a real space chiral spin texture. Using electrical magnetoresistance measurements at low temperature, we observe an emergent excess contribution to the transverse magneto-resistance whose behavior is consistent with a topological Hall effect arising from the formation of an interfacial chiral spin texture. This excess Hall voltage varies with gate voltage, indicating a promising electrostatically-tunable platform for understanding the interplay between the helical momentum space states of a Dirac semimetal and chiral real space spin textures in a ferromagnet.

\end{abstract}

\maketitle



\section*{Introduction}
Topologically non-trivial spin textures in real space (such as skyrmions, antiskyrmions, and merons) continue to attract significant scientific interest in magnetic materials because they provide a platform for fundamental studies of emergent electrodynamic phenomena and due to their potential applications for energy-efficient, high density information storage~\cite{Fert_RevModPhys.96.015005,wang2022fundamental,dohi2022thin,GOBEL20211,nagaosa2013topological,,kimbell2022challenges,yang2021chiral}. These real space spin textures arise from a complex interplay of several factors such as the anisotropic Dzyaloshinskii–Moriya (DM) interaction, long-range dipolar interaction, frustrated exchange interactions, magnetic anisotropy, and external applied magnetic field. Various types of magnetic spin textures have been reported, including Bloch-type skyrmions induced by a bulk DM interaction~\cite{wang2020giant}, Néel-type skyrmions primarily linked to interfacial DM interaction~\cite{wu2020neel}, and spin chirality arising from geometrical frustration due to the interplay of different magnetic interactions~\cite{dzyaloshinskii1957ie,moriya1960anisotropic,bogdanov2001chiral,wang2022topological}.
Quantum materials with a topological band structure have played an important role in this general context because the strong spin-orbit coupling (SOC) that gives rise to the underlying band inversion and momentum-space Berry curvature can also influence the real space behavior of spins. Indeed, evidence for chiral magnetic spin textures has been found in topological insulator heterostructures \cite{li2020topological,jiang2020concurrence,xiao2021mapping} and in Weyl semimetal thin films \cite{Taylor_PhysRevB.101.094404}.

Expanding the landscape of topological quantum materials that can host chiral magnetic spin textures is thus valuable, especially if one can design a platform that allows external tuning of the underlying topological band structure using electric and/or magnetic field. The archetypal Dirac semimetal (DSM) \CdAs, a three-dimensional analog of graphene, offers an attractive opportunity in this regard. This DSM has a simpler Dirac node structure compared with many common Weyl semimetals (such as TaAs or Mn$_3$Sn). Epitaxial growth of \CdAs~thin films allows electrostatic control over the chemical potential \cite{Galletti_PhysRevB.97.115132, Xiao_PhysRevB.106.L201101}, while interfacing with a ferromagnetic material (in principle) allows the breaking of time-reversal symmetry and introduction of anisotropic exchange interactions \cite{Xiao_PhysRevMaterials.6.024203,Mitra_PhysRevMaterials.7.094201}. Additionally, in contrast to the low mobility metallic magnetic systems that have been thus far used for studying topologically non-trivial chiral spin textures, \CdAs~thin films have a relatively high carrier mobility~\cite{schumann2016molecular,schumann2018observation,nishihaya2019quantized, Xiao_PhysRevB.106.L201101}, even when interfaced with a ferromagnet~\cite{Xiao_PhysRevMaterials.6.024203}; this provides a strong motivation for realizing skyrmions and other chiral spin textures in a DSM such as \CdAs.

Here, we use a concerted theory-experiment effort to identify a promising DSM-based heterostructure, \CdAs/\InMnAs, that allows for electrostatically controlled chiral spin textures. We choose this combination of a DSM and a ferromagnetic semiconductor for three primary reasons. First, we find that we can synthesize \CdAs/\InMnAs~heterostructures with a chemically sharp interface. Second, the epitaxial growth of \InMnAs~on a standard III-V substrate leads to ferromagnetic order with perpendicular magnetic anisotropy (PMA)~\cite{Oiwa_PhysRevB.59.5826,ohno2000electric,Steinebronn_2024}, thus enhancing the possibility of perturbing the helical Dirac states in the DSM via broken time-reversal symmetry. Third, we can choose the composition of \InMnAs~to have a resistivity in the range of $\sim 10$ m$\Omega$.cm \cite{Oiwa_PhysRevB.59.5826}, about an order of magnitude greater than that of \CdAs~($\sim 1$ m$\Omega$.cm \cite{Xiao_PhysRevB.106.L201101}), while still hosting robust ferromagnetic order. This allows convenient transport probes of the interplay between the Dirac states and local moments. Our calculations indicate a nonzero off-diagonal spin susceptibility in such heterostructures due to Rashba SOC from broken inversion symmetry, implying the presence of an interfacial DM interaction between the Mn spins, mediated by the Dirac electrons in the DSM. This can lead to the formation of chiral spin texture of these local moments. We use electrical magnetoresistance measurements to identify necessary (but not yet sufficient) signatures of a gate-voltage-tunable topological Hall (TH) effect, consistent with the presence of chiral spin textures implied by our calculations.  

Evidence for chiral spin textures in quantum materials can be obtained either directly via neutron diffraction~\cite{mühlbauer2009skyrmion}, Lorentz microscopy~\cite{yu2010real,uchida2006real}, x-ray microscopy~\cite{litzius2017skyrmion}, and magnetic force microscopy~\cite{soumyanarayanan2017tunable,raju2019evolution} or indirectly via electrical transport~\cite{li2020topological}.
In the latter case, spin texture is manifested as an excess Hall resistance near the magnetic coercive field, denoted as the TH effect. In contrast to the intrinsic anomalous Hall effect (AHE), which is related to the Berry curvature in momentum space, the TH effect is related to the accumulation of Berry phase in real space. Consistent with the implications of our calculations, our experiments show an emergent TH effect at low temperature ($T \lesssim 4$~K), characterized by a peak in the Hall resistance near the coercive field. We map out the behavior of the TH effect as a function of temperature, magnetic field, and gate voltage in electrostatically top-gated devices, showing that its signature is enhanced as we approach the hole doped regime in \CdAs. This strongly suggests the formation of a Dirac-electron-mediated chiral spin texture at the DSM/ferromagnet interface.

\section*{Results}
\subsection*{Theory: from band structure to spin susceptibility and DM interaction}

\begin{figure}[htbp]
    \includegraphics[width=14cm]{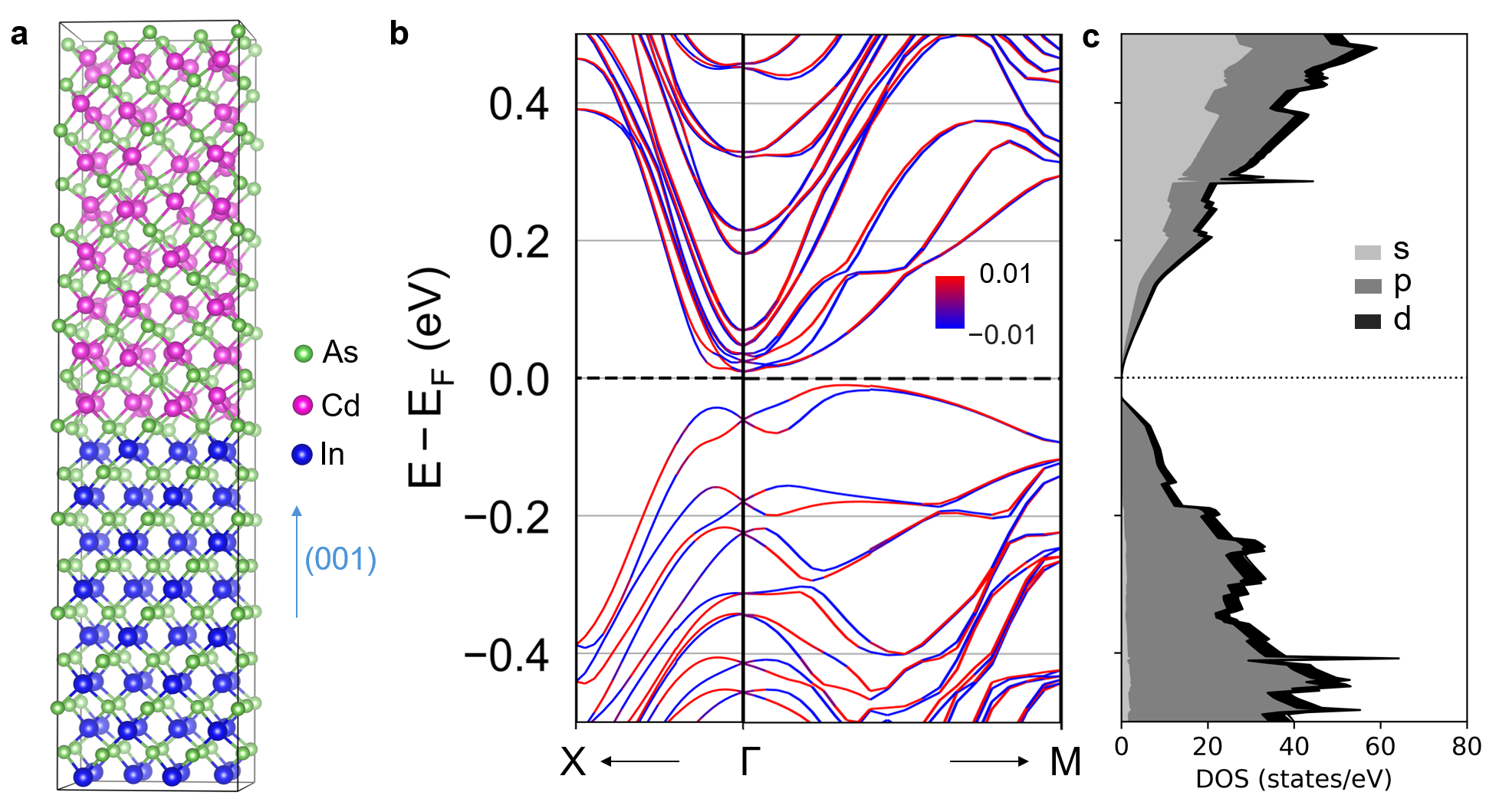}
    \caption{{\bf{Band structure of Cd$_3$As$_2$/InAs heterostructure.}} {\bf a} Side view of a sixteen-layered Cd$_3$As$_2$/InAs heterostructure. {\bf b} Band structure near Fermi-energy (E$_F$) with SOC based on the DFT calculations. Red and blue colored lines represent positive and negative spin polarization along the y direction. The color bar represents the weight of the spin. {\bf c} Orbital projected DOS from the DFT calculation. Shading indicates the cumulative contributions of the projected components of the DOS. The inset shows the colors for s,p, and d orbitals.} 
    \label{Fig1}
\end{figure}

We begin by discussing whether a DM interaction can emerge from the electronic band structure of a \CdAs/\InMnAs~heterostructure as a result of the interplay between the SOC and the crystal symmetry. We show this starting with a first-principles electronic structure of the heterostructure at the density functional theory (DFT) level, followed by parameterizing a low-energy effective Hamiltonian to calculate spin susceptibilities (see methods for details on DFT setup). 
For clarity of the electronic structure to be calculated, no Mn dopants are explicitly included in InAs for our heterostructure models introduced below, since they are expected to only introduce local moments without significantly perturbing the InAs electronic structure or contribute to spin susceptibilities that are dominated by \CdAs. To verify this, we inspected bands near the Fermi level of InAs (no \CdAs~yet) Mn-doped at a concentration of 1.6 $\%$ from spin-polarized calculations. Band edge dispersions and orbital characters remained almost the same except for spin splittings in the valence bands (See Supplementary Section I). The heterostructure model, shown in Fig.~\ref{Fig1}(a), is constructed by connecting a $2\times2\times4$ supercell of conventional InAs unit cell with a conventional unit cell of Cd$_3$As$_2$ along the (001) direction. A $5.3 \%$ in-plane compressive strain is applied to the Cd$_3$As$_2$ slab, since we find that the Dirac point energy of Cd$_3$As$_2$ is less susceptible to strain then the band edge energies of InAs. 

The DFT band structure including spin-orbit coupling for the heterostructure is shown in Fig.~\ref{Fig1} (b). Near the Fermi energy, both the conduction and valence bands mainly come from the Cd$_3$As$_2$ layer due to its semi-metallic nature. A salient feature of the band dispersion in Fig. \ref{Fig1} (b) is that most valence bands show a strong spin splitting with Rashba character, while the spin splitting of most conduction bands is much weaker. We attribute this feature to different atomic orbital characters. The atomic orbital projected density of states (DOS) in Fig. \ref{Fig1}(c) shows that the valence bands are dominated by arsenic p-orbitals, which can host a strong atomic SOC, while the conduction bands mainly originate from cadmium s-orbitals with no atomic SOC. 


\begin{figure}[htbp]
    \includegraphics[width=15cm]{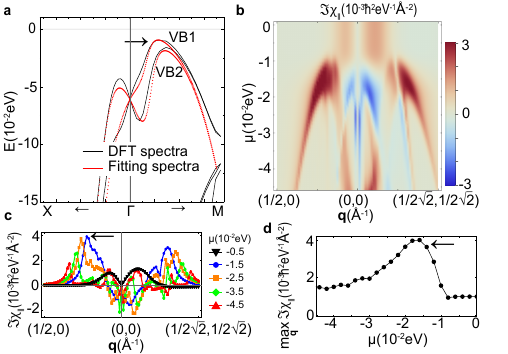}
    \caption{ {\bf{Theoretical calculation of spin susceptibilities and anomalous Hall conductivity.}}
    {\bf a} The energy spectra of the model Hamiltonian in Eq.~(\ref{eq:four band Hamiltonian}) (red dashed lines) fitted from the DFT spectra (black solid lines) for the valence bands. The black arrow points to the valence band top at the $\Gamma$-M line. VB1, VB2 are the top two valence bands.
    {\bf b} Spin susceptibility $\Im \chi_\parallel(\vec q,\mu)$.
    {\bf c} $\Im \chi_\parallel(\vec q,\mu)$ for $\mu = -0.005, -0.015, -0.025, -0.035, -0.045$~eV. The black arrow points to the maximum of $\Im \chi_\parallel(\vec q,\mu = -0.015$eV$)$ along $\vec q = (q_x,0)$.
    {\bf d} The maximum spin susceptibility $\max_{\vec q} \Im \chi_\parallel(\vec q,\mu)$ along $\vec q = (q_x,0)$ under different $\mu$. The black arrow in (d) points to $\max_{\vec q} \Im \chi_\parallel(\vec q,\mu = -0.015$~eV$)$ corresponding to the peak indicated by the black arrow in (c).
    } 
    \vspace{3mm}
    \label{Fig2}
\end{figure}


Since the DM interaction strength is proportional to the off-diagonal components of the spin susceptibility~\cite{jiang2020concurrence}, we compute the spin susceptibility of the \CdAs/\InMnAs~ heterostructure, focusing
on the top valence bands with a large spin splitting. Although the spin splitting of the valence band resembles the Rashba type, the strong anisotropy of the valence band dispersion along the $\Gamma-X$ and $\Gamma-M$ directions cannot be described by a simple Rashba model (see Supplementary Section V). 
To fully account for the band anisotropy, we include a set of s-orbital conduction bands in addition to the p-orbital valence band, and develop a 4-band model based on the $C_{4v}$ group (the little group at $\Gamma$ for the heterostructure). It should be noted that this 4-band model can only describe the dispersion of valence bands, but not the conduction bands. 
We take the basis of the Hamiltonian as $\vert p, J_z = \pm 1/2 \rangle$ and $\vert s, J_z = \pm 1/2 \rangle$ with two atomic orbitals $s,p$ for conduction and valence bands, respectively, where $J_z$ is the total angular momentum under the four-fold rotation. The Hamiltonian reads 
\begin{equation}
    H (\vec k) =
    \begin{pmatrix}
    E_p + m &  i A k_- & V_{sp} & 0 \\
    - i A k_+ & E_p - m & 0 & V_{sp}\\
    V_{sp} & 0 & E_s + m & 0\\
    0 & V_{sp} & 0 & E_s - m
    \end{pmatrix},
    \label{eq:four band Hamiltonian}
\end{equation}
where $k_{\pm} = k_x \pm i k_y$, $E_p = E_{p0}+B_p (k_x^2 + k_y^2 ) + D_{p} k_x^2 k_y^2$, $E_s=E_{s0}$, the hybridization between $s,p$ orbitals $V_{sp} = B_{sp} (k_x^2 + k_y^2 ) + D_{sp} (k_x^2+k_y^2)^2$ and the magnetization $m$.
The $Ak_\pm$ is the Rashba spin splitting term in the valence p-orbital bands, while such a term is absent in the s-orbital conduction band.
The $D_p k_x^2 k_y^2$ term in $E_p$ captures the band dispersion anisotropy in the DFT spectra.
The dispersion of two valence bands, VB1 and VB2 in Fig.~\ref{Fig2}(a), fit well with the DFT bands. The material dependent  parameters are given by $A = 0.4$~eV\r{A}$^{-1}$, $E_{p0} = - 0.061$~eV, $B_p = -3$~eV\r{A}$^{-2}$, $D_{p} = 7000$~eV\r{A}$^{-4}$, $E_{s0} = 0.024$~eV, $B_{sp}= 6$~eV\r{A}$^{-2}$, and $D_{sp}= 80$~eV\r{A}$^{-4}$.
The spin susceptibility is then evaluated by the linear response formula~\cite{mahan2000many}
\begin{equation}
\begin{split}
    \chi_{\alpha \beta} (\vec q, \mu) = & \int \frac{d^2 \vec k}{(2\pi)^2} \sum_{n^\prime,n} 
    \frac{n_{\text F}(\epsilon_{n^\prime}(\vec k))-n_{\text F}(\epsilon_{n}(\vec k + \vec q))}{i\Gamma+ \epsilon_{n^\prime}(\vec k) - \epsilon_{n}(\vec k + \vec q)}  \\
    & \text {Tr} \Big( P_{n^\prime}(\vec k) (\frac{s_\alpha}{2}) P_n(\vec k + \vec q) (\frac{s_\beta}{2} ) \Big),
\end{split}
\end{equation}
where $n^\prime, n$ are band indices, $\alpha, \beta = x, y, z$, $\epsilon_{n}(\vec k)$ is the eigen-energy of the model Hamiltonian Eq.(\ref{eq:four band Hamiltonian}), $P_n(\vec k) = \vert \vec k,n \rangle\langle \vec k,n\vert$ is the projection operator to the eigenstates $\vert \vec k,n \rangle$, $\mu$ is the chemical potential, $\Gamma$ is the self-energy broadening from disorder and $n_{\text F}(\epsilon)=1/(1+e^{(\epsilon-\mu)/k_\text B T})$ is the Fermi-Dirac distribution. We choose $\Gamma = 1$~meV and $k_\text B T=1~$meV for the calculation in \figref{Fig2}(b). $s_\alpha = \tau_0 \sigma_\alpha$ is the spin operator with $\tau,\sigma$ are Pauli matrices for the orbital, spin degrees of freedom, respectively.
We choose the magnetization term $m = 0$~eV for the calculation of spin susceptibility. The DM interaction strength is given by
$D_\gamma (\vec q, \mu) = \varepsilon_{\gamma\alpha\beta} \chi_{\alpha\beta}  (\vec q, \mu)$\cite{jiang2020concurrence}, where $\varepsilon$ is the Levi-Civita symbol . For three off-diagonal spin susceptibility $\chi_{xy} (\vec q, \mu)$, $\chi_{xz} (\vec q, \mu)$, $\chi_{yz} (\vec q, \mu)$ that can contribute to the DM interaction, the time reversal symmetry $\hat{\mathcal T}$ requires $\chi_{\alpha\beta}(\vec q, \mu) = \chi_{\alpha\beta}(-\vec q, \mu)^*$ and the two-fold rotation about z-axis $\hat{\mathcal C}_{2z}$, together with $\hat{\mathcal T}$, requires $\chi_{xz} (\vec q, \mu)$, $\chi_{yz} (\vec q, \mu)$ to be purely imaginary and $\chi_{xy} (\vec q, \mu)$ to be purely real.
On the high symmetry lines X-$\Gamma$-M, the in-plane mirror symmetries, $\hat{\mathcal M}_{010}$ at $\Gamma$-X and $\hat{\mathcal M}_{\bar{1}10}$ at $\Gamma$-M,
give rise to the constraint for non-zero off-diagonal spin susceptibility $\chi_\parallel (\vec q,\mu)$, which is defined as
\begin{equation}
    \chi_\parallel (\vec q,\mu) = \chi_{xz}(\vec q, \mu) q_x/ \vert \vec q \vert + \chi_{yz}(\vec q, \mu) q_y/ \vert \vec q \vert.
\end{equation}
(See more details in the Supplementary Section III). Figure~\ref{Fig2}(b) shows the spin susceptibility,  $\Im\chi_\parallel(\vec q, \mu)$, as a function of $\mu$ and $\vec q$ along X-$\Gamma$-M.
In \figref{Fig2}(c), we show the spin susceptibility at fixed chemical potentials.
The spin susceptibility $\Im \chi_\parallel((q,0), \mu)$ for $\vec q$ along $\Gamma$-X is different in magnitude from $\Im \chi_\parallel((q/\sqrt 2,q/\sqrt 2), \mu)$ for $\vec q$ along $\Gamma$-M, showing the anisotropy in spin susceptibility (See comparisons with isotropic Rashba model in the Supplementary V).
The peak values of the spin susceptibility along the direction $\vec q = (q_x,0)$ is defined as $\max_{\vec q} \Im \chi_\parallel(\vec q, \mu)$.
$\max_{\vec q} \Im \chi_\parallel(\vec q, \mu)$ for $\mu = -0.015$~eV is depicted by the black arrow in \figref{Fig2}(c).
$\max_{\vec q} \Im \chi_\parallel(\vec q, \mu)$ is plotted in \figref{Fig2}(d) for different chemical potentials $\mu$.
We note that the maximum spin susceptibility increases rapidly as the chemical potential is lowered towards the valence bands, corresponding to an increase of DM interaction, as well as the TH effect, in the valence bands. It also reveals a maximum near the valence band top and then decreases when the chemical potential is further lowered.

Since the heterostructure model calculation above is carried out in a single unit cell thick \CdAs~quantum well structure ($\sim$2.7~nm thickness) for computational considerations, it is important to comment on the implications for heterostructures with greater thicknesses (in the range 10 - 20 nm), as typically encountered in experimental studies \cite{Galletti_PhysRevB.97.115132,Xiao_PhysRevB.106.L201101}. The large tunability of the DM interaction and the TH effect derives from the separation of s- and p-orbital character in the conduction and valence bands, where only p-orbitals have a strong atomic SOC and thus can support large Rashba SOC splittings when inversion symmetry is broken near the interface. As we go beyond the regime of extreme quantum confinement addressed in our calculations, more quantum well subbands will crowd near the Fermi energy~\cite{PhysRevB.88.125427}, reducing the gap to a few meV, until a Dirac cone is restored in the 3D bulk limit. Since the s- and p-orbital character separation in the two band edges persists in the bulk limit (see Supplementary Section IV), we expect that the strong SOC still exists for the valence bands in intermediate thicknesses (e.g. $25$~nm). Furthermore, we expect the formation of quantum well subbands near the \CdAs/\InMnAs~heterointerface due to the band bending from the built-in electric field induced by interlayer charge transfer. Since the Rashba SOC requires the breaking of inversion symmetry, we anticipate that only the quantum well sub-bands confined closer to the interface can give rise to the DM interaction and the TH effect. 

\subsection*{Experiment: synthesis/characterization of heterostructures and electrical magnetoresistance measurements}

We now discuss experiments carried out to examine the theoretical picture presented earlier. We use molecular beam epitaxy (MBE) to synthesize \CdAs/\InMnAs~heterostructures on (001) GaAs substrates after the growth of a thin layer of $2$~nm GaAs followed by a buffer layer of $50$~nm GaSb. The GaSb buffer serves to relax the $\sim 7\%$ lattice mismatch with the GaAs substrate through the formation of misfit dislocations. The detailed growth conditions are described in Methods. As well-established in past studies \cite{Oiwa_PhysRevB.59.5826}, the epitaxial growth of \InMnAs~is carried out at a low substrate temperature to avoid the formation of MnAs clusters. The chosen In/Mn flux ratio is intended to yield an \InMnAs~composition in the range $x \sim 0.02 - 0.03$. The epitaxy of \CdAs~on top of the \InMnAs~layer is carried out using growth conditions similar to those reported in the literature for growth on (001) GaAs, albeit with a different buffer layer structure \cite{kealhofer2021thickness}. 

\begin{figure}[htbp]
   \centering
    \includegraphics[width=16cm]{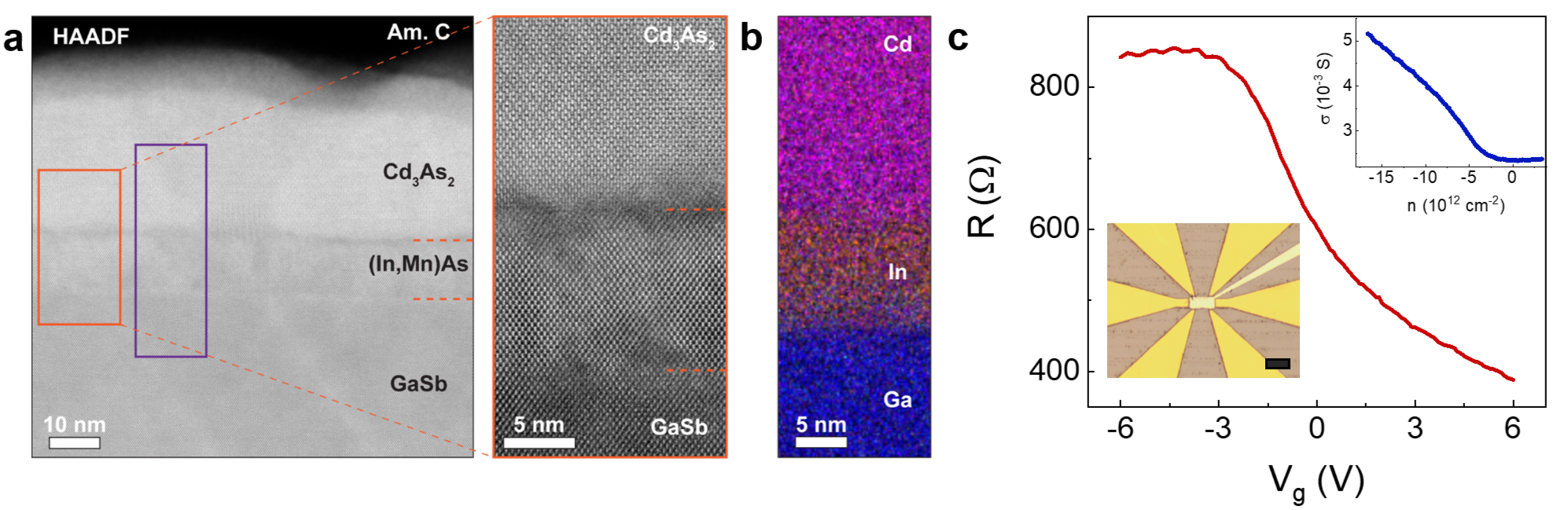}
    \caption{ {\bf{Heterostructure characterization and device characteristics.}} {\bf a} Cross-sectional HAADF-STEM image of a Cd$_3$As$_2$/\InMnAs~ heterostructure. Atomic-resolution image from the region highlighted in the orange box, demonstrating the orientation between the different layers, showing the [111] orientation in the  Cd$_3$As$_2$ layer. {\bf b} STEM-EDX maps from the region in the purple box in (a) showing elemental distribution of Cd, In and Ga in the heterostructure layers. {\bf c} Resistance ($R$) vs gate-voltage ($V_g$), showing charge neutrality point at $V_g=-4$.  The upper right inset shows conductivity ($\sigma$) vs. carrier density ($n$). The lower left inset shows an optical micrograph of a typical device. The scale bar is $100$~$\mu$m.} 
    \vspace{3mm}
    \label{Fig3}
\end{figure}

Fig.~\ref{Fig3}(a) shows a high-angle annular dark-field scanning transmission electron microscope (HAADF-STEM) image of the heterostructure studied in this work. Atomic-resolution imaging at the heterostructure interfaces shows a smooth interface between the GaSb-\InMnAs~ and the Cd$_3$As$_2$ layers, with grain boundaries visible in each layer. High-resolution STEM data shows the formation of a coherent crystalline heterostructure, albeit with some misfit dislocations that thread through the entire heterostructure \cite{Steinebronn_2024}. This is also seen in the compositional distribution of the elements across the heterostructure which is shown in Fig.~\ref{Fig3}(b). The thickness of the \InMnAs~and \CdAs~layers is $17$~nm and $25$~nm, respectively. Measurements of control samples consisting of \CdAs/GaSb/GaAs (001) and \InMnAs/GaSb/GaAs (001) \cite{Steinebronn_2024} show that the resistivity of the \InMnAs~($\approx5$~k$\Omega$/sq) layer is one order of magnitude higher than Cd$_3$As$_2$~($\approx570$~$\Omega$/sq at $V_g=0$~V)~\cite{Steinebronn_2024}. Thus, we expect that in our (17 nm) \InMnAs/(25 nm)\CdAs~heterostructure devices, the current flows primarily in the \CdAs~layer. The longitudinal resistance ($R$) as a function of gate-voltage ($V_g$) in Dev I is shown in Fig.~\ref{Fig3}(c), where we observe a charge neutrality point (CNP) around $V_g=-4$~V, indicating the films to be electron-doped. The field effect mobility ($\mu$) of these devices, calculated using $\mu=\sigma/ne$ ($\sigma$ and $n$ are conductivity and carrier density respectively) is typically $\approx 1100$~cm$^2$/Vs (inset of {Fig.~\ref{Fig3}(c))}.

\begin{figure}[htbp]
    \includegraphics[width=12cm]{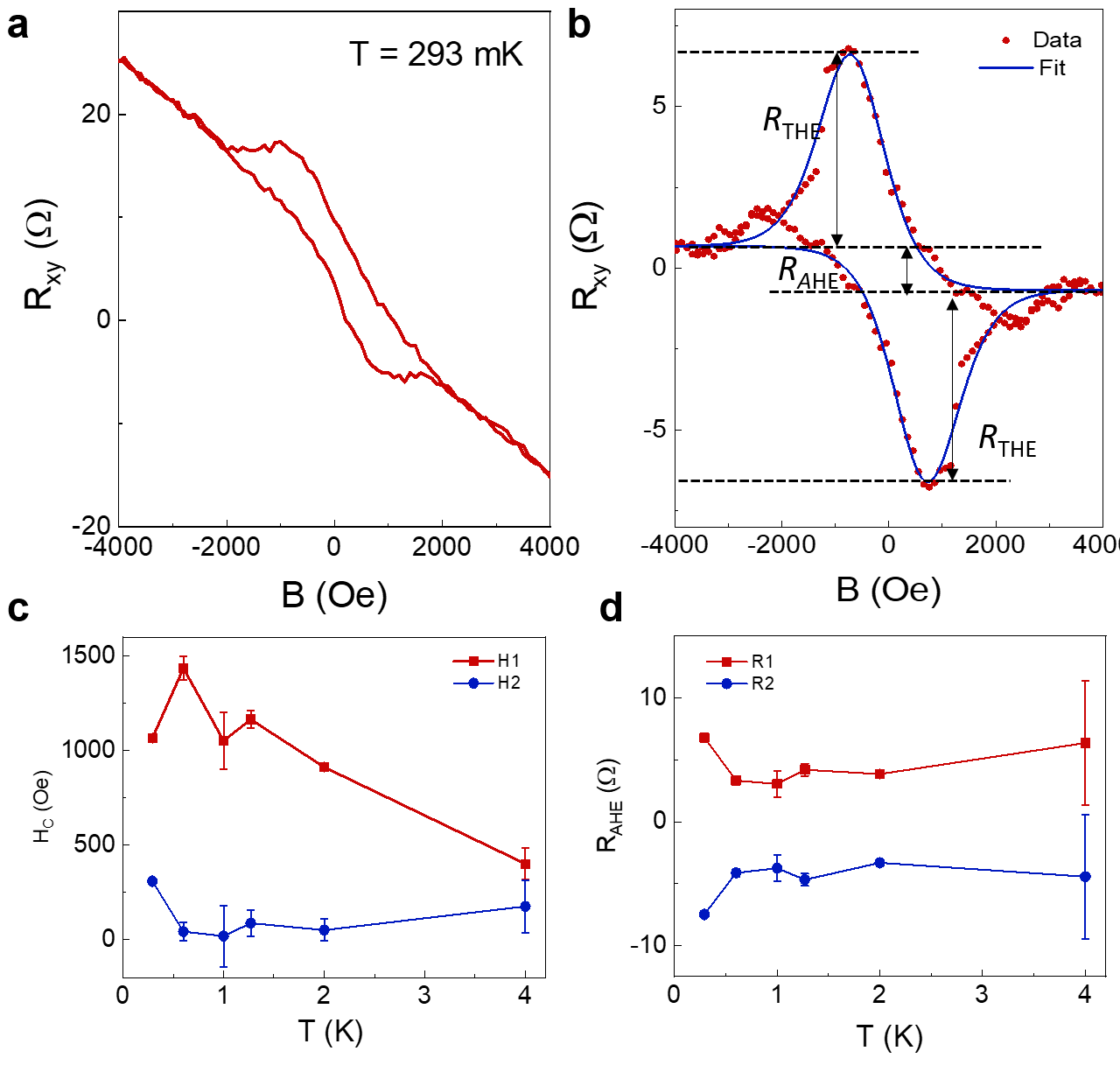}
    \caption{{\bf Measurement of anomalous and TH effect.} {\bf a} Measurement of the Hall resistance ($R_{xy}$) versus magnetic field in Dev1 at $T=293$~mK at $V_g=0$~V). {\bf b} Hall resistance in Dev 1 after subtracting the linear in magnetic field ordinary Hall effect contribution at $T=293$~mK and $V_g=0$~V. The solid line is a fit to the data using Eq. 5. $R_{THE}$ and $R_{AHE}$ denote the magnitude of the TH effect and AHE, respectively. {\bf c} and {\bf d} The temperature dependence of the coercive field ($H_{c1,2}$) and the anomalous Hall resistance ($R_{AHE1,2}$), respectively, extracted by fitting the data using the Eq. 5.} 
    \vspace{3mm}
    \label{Fig4}
\end{figure}

We carry out electrical magnetoresistance measurements on lithographically patterned Hall bars (See Methods). The measurements in this manuscript focus on two top-gated Hall bar devices (Dev 1 and Dev 2) fabricated from a single wafer. The transverse magnetoresistance data for Dev 1 at $T=293$~mK and $V_g=0$V is provided in Fig.~\ref{Fig4}(a). The total Hall resistance $R_{xy}$ is the combination of the ordinary Hall effect ($R_{OHE}$), the AHE ($R_{AHE}$), and the 
TH effect ($R_{THE}$) given by:
\begin{equation}
R_{xy}=R_{OHE}+R_{AHE}+R_{THE}=\frac{R_o}{d}B+\frac{R_s}{d}M+R_{THE} 
\end{equation}
where $R_0$, $R_s$, $B$, $M$, and $d$ are the ordinary Hall coefficient, AHE coefficient, magnetic field, magnetization perpendicular to the layer, and the thickness of the film respectively. The Hall data shown in Fig.~\ref{Fig3}(b) has been anti-symmetrized ($R_{xy}=\frac{R(B)-R(-B)}{2}$), and we have subtracted the linear ordinary Hall component. 
The excess voltage due to the TH effect manifests as a distinct peak in the Hall resistivity, shown in Fig.~\ref{Fig4}(b) (see Supplementary Section VI). Note that the AHE in control samples of \InMnAs~only shows a standard AHE\cite{Steinebronn_2024}. We are aware that peaks attributed to the TH effect can arise due to the co-existence of two or more overlapping AHE contributions of opposite sign \cite{Laurens_PhysRevX.10.011012,li2020topological,tai2022distinguishing}. To eliminate this scenario, we use temperature-dependent transverse magneto-resistance measurements. In a recent analysis~\cite{tai2022distinguishing}, THE effect artifact peaks arising from a two-component AHE contribution have been shown to result in a reversal of the polarity of the Hall loops with temperature and gate-voltage; this behavior which is absent in our samples. We also carry out a more detailed analysis by
fitting our Hall data to the following equation, consisting of two AHE components: 
\begin{equation}
\begin{split}
R_{AHE}(B)=R_{AHE1}\tanh\left( \frac{B \pm H_{c1}}{H_{01}} \right)+
\\ R_{AHE2}\tanh\left( \frac{B \pm H_{c2}}{H_{02}} \right)
\label{5}
\end{split}
\end{equation}

Here, the first and second terms on the right side denote the two anomalous Hall components, $R_{AHE1}$ ($R_{AHE2}$) and $H_{c1}$ ($H_{c2}$) are the amplitude and coercivity of the  two components, 
while $H_{01}$ ($H_{02}$) are fitting constants. The $T$-dependence of the extracted parameters $H_{c1,2}$ and $R_{AHE1,2}$ is shown in Fig.~\ref{Fig4}(c,d) respectively. Although we find a reasonable fit to the experimental data using Eq. 5 (Fig.~\ref{Fig4}(b)), the extracted values do not show a physically meaningful $T$-dependence: the amplitude of the AHE increases with decreasing temperature for the two contributions of opposite sign, while the corresponding coercive field increases for one of the components, but decreases for the other. This leads us to the conclusion that the observed excess Hall voltage in unlikely to arise from the superposition of two AHE signals with opposite signs. 

\begin{figure}[htbp]
    \includegraphics[width=16cm]{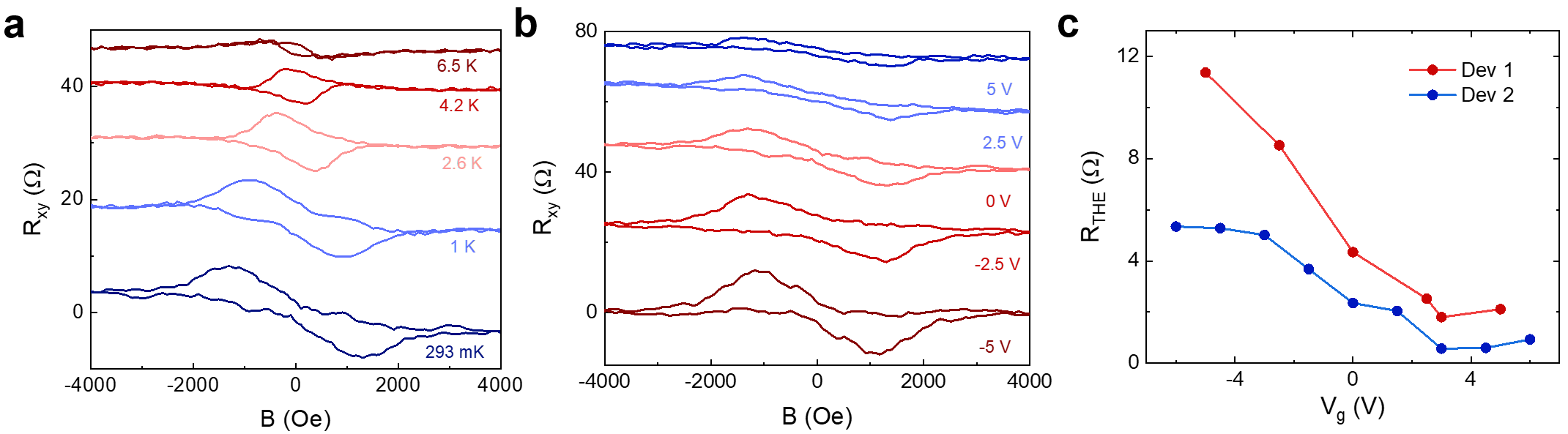}
    \caption{{\bf{Temperature and gate-voltage dependence of AHE.}} {\bf a} Plot showing $R_{xy}$ after subtracting linear ordinary Hall component as a function of  perpendicular field ($B$) at different temperatures ($T$) in Dev 2 at $V_g=0$. The curves have been offset for clarity. {\bf b} Hall resistance after subtracting linear ordinary Hall component as a function of gate-voltage ($V_g$) at $T=293$~mK. The curves have been offset for clarity. {\bf c} Gate-voltage dependence of the magnitude of the TH effect, $R_{THE}$, showing an increase as $V_g$ is towards the hole-doped regime.} 
    \vspace{1mm}
    \label{Fig5}
\end{figure}

For further insight into the origin of the excess Hall effect, we map out the temperature and gate-voltage dependence of the Hall resistance. 
We observe a few key features of the $T$-dependence of $R_{xy}$. First, the excess Hall resistance is present only in a specific range of $B$. Usually, spin texture stabilization occurs within a small range of magnetic field that depends on the strength of the magnetic free energy and the external field. This is consistent with the behavior in Fig.~\ref{Fig5}(a), where we observe that the peak is strongest between $1000$ to $2000$~Oe for Dev1, and the range gradually diminishes as $T$ increases.
Second, the excess Hall resistance is present only below a critical temperature lower than the ferromagnetic Curie temperature of the sample. Fig.~\ref{Fig5}(a) indeed shows that the peak in $R_{xy}$ disappears above $T\approx 4.2$~K, whereas the sample is still ferromagnetic upto somewhat higher temperature. Third, the strength of the TH effect, given by $R_{THE}$, reduces as $T$ is increased towards $T_{c}$ (see Supplementary Section VII). 

We also measure the Hall effect as a function of $V_g$ to investigate the effect of changing carrier density on the observed behavior of the TH effect. Fig. \ref{Fig5}(b) and 5(c) show that the peak in $R_{xy}$ and the magnitude of $R_{THE}$ increases with increased hole doping in the system as $V_g$ is tuned towards more negative values, indicating that Dirac electrons play a significant role in the observed effect.

\section*{Discussion}

Our magneto-transport data shows a strong TH effect in the hole doping regime, with a significant reduction in the electron doping regime. This qualitatively coincides with the strong SOC-induced spin splitting for the valence bands and negligible spin splitting of most conduction bands. A large spin susceptibility is also found when the chemical potential is tuned to the valence bands VB1 and VB2 below $\mu=-0.01$ eV (black arrow in \figref{Fig2}(b)) from the band gap, indicating a strong DM interaction in the valence bands. This is consistent with the experiments that a large THE is observed in the hole-doped regime.
Indeed, it has been well established that the DM interaction \cite{dzyaloshinskii1957ie,moriya1960anisotropic} that originates from Rashba type of SOC can lead to chiral spin textures (e.g. skrymions and chiral magnetic domain walls) to give rise to the TH effect \cite{nagaosa2013topological,neubauer2009topological,lee2009unusual,kanazawa2011large,huang2012extended,hou2017thermally,bottcher2018b,wang2019spin,yasuda2016geometric,jiang2020concurrence}. 
The presence of strong DM interaction also explains the temperature dependence of the excess Hall resistance shown (See Supplementary Section VII) which arises from the $T$-dependence of the DM interaction constant $D$. Spin texture such as skyrmions are formed when $D\ge\frac{4}{\pi}\sqrt{AK}$, where $A$ is the exchange constant and $K$ is the perpendicular anisotropy constant. The $T$-dependence is related to the interplay between $A$ and $K$. When $T$ increases, both $D$ or $A$ and $K$ decrease, but the $D$ or $A$ can have a stronger $T$-dependence than $K$, leading to a violation of the equality. The magnitude of $R_{THE}$ is given
as~\cite{bruno2004topological,matsuno2016interface,he2018exchange,li2020topological}: 
\begin{equation}
R_{THE}=\frac{P}{en_{2D}}b=\frac{P}{en_{2D}}\frac{\Phi_0}{d^2} 
\end{equation}
Here $P$, $n_{2D}$, $e$, $\Phi_0$ and $d$ are the spin polarization rate in Cd$_3$As$_2$, electron
charge,  sheet carrier density in Cd$_3$As$_2$, magnetic flux quantum, and the average distance between centers of two neighboring skyrmions.
which scales as $\frac{2\pi A}{D}$~\cite{banerjee2014enhanced}. Both $d$ and $n_{2D}$ increase with $T$, which leads to a reduction in $R_{THE}$~\cite{li2019magnetization}. 

To gain further insight into the experimental observations (Fig. \ref{Fig4}d), we carry out additional calculations of the anomalous Hall conductivity $\sigma_{xy}(\mu)$ for the valence band in Fig.~\ref{Fig2}(e); this is induced by the overall ferromagnetism in the heterostructure.
To induce AHE, we include an exchange coupling term with nonzero $m$ in the model Hamiltonian Eq.(\ref{eq:four band Hamiltonian}). The anomalous Hall conductivity is then calculated from 
\begin{equation}
\begin{split}
    \sigma_{xy} (\mu) = &\int \frac{d^2 \vec k}{(2\pi)^2} \sum_{n^\prime,n}\frac{n_{\text F}(\epsilon_{n^\prime}(\vec k))-n_{\text F}(\epsilon_{n}(\vec k))}{i\Gamma+ \epsilon_{n^\prime}(\vec k) - \epsilon_{n}(\vec k)} \\ &\text {Tr} \Big( P_{n^\prime}(\vec k) v_x P_n(\vec k) v_y \Big).
\end{split}
\end{equation}
$v_x = \partial_{k_x}H(\vec k), v_y = \partial_{k_y}H(\vec k)$ are the velocity operators.
When the chemical potential is tuned to the valence bands, the anomalous Hall conductivity also shows a peak due to the large Berry curvature from the band inversion between $s,p$ orbitals (see details in Supplementary Section IV). 


In conclusion, the prediction of a non-zero spin susceptibility in \CdAs/\InMnAs~heterostructures and the observation of the TH effect in such samples provides strong evidence of the presence of spin texture, although its exact nature, such as the presence of chiral domain wall or skyrmions, remains to be understood using probes other than electrical transport. The maximum temperature for formation of spin texture can also be increased by growing thin films on suitable ferromagnetic materials with higher Curie temperatures. Overall, our work provides a new high mobility material platform for observing non-trivial chiral spin texture, and motivates further theoretical and experimental investigation.

\section*{Methods}

\subsection*{Density functional theory calculations}

The band structure was calculated using density-functional theory (DFT) as implemented in VASP~\cite{kresse1996efficient}, using the Perdew-Burke-Ernzerhof (PBE) exchange-correlation functional~\cite{perdew1996generalized} and projector augmented wave (PAW) pseudo-potentials~\cite{blochl1994projector}. The cut-off for the kinetic energy of plane waves is set to $400$~eV. The heterostructure is optimized until the total residual energy and force on each atom were less than $10^{-5}~$eV and $0.1$~eV/Å. A $\Gamma$-centered $k$-point mesh of $2 \times 2 \times 1$ is used to optimize the structure. All atoms and the lattice parameter along the $c$ direction are fully relaxed. Furthermore, we use the Hubbard $U$~\cite{anisimov1991band} parameters for the InAs  ($U_{eff}^{In,5p}=-0.5$~eV  and $U_{eff}^{As,4p}=-7.5$~eV) slab in the heterostructure from Ref.~\cite{yu2020machine}. This choice of $U$ parameters was reported to yield a band gap that agrees well with experimental values for bulk InAs. 

\subsection*{MBE growth of \CdAs/\InMnAs~heterostructures}

We synthesize \CdAs/\InMnAs~heterostructures on (001) GaAs substrates using molecular beam epitaxy (MBE) in a Veeco 930 chamber. The source materials are evaporated from standard effusion cells containing Ga (99.999995\%), As (99.99999\%), Sb(99.9999\%), Mn (99.9998\%), and a high purity compound source of Cd$_3$As$_2$. The native oxide on the GaAs substrate is initially desorbed at a substrate temperature ($T_s$) of $600$~$^\circ$C until the RHEED becomes streaky. We then grow a thin buffer layer of $2$~nm GaAs with a As/Ga beam equivalent pressure ratio (BEPR) of 17, followed by an additional buffer layer of $50$~nm GaSb at $T_s=440$~$^\circ$C (BEPR of Sb/Ga=7). The GaSb buffer serves to relax the $\sim 7\%$ lattice mismatch with the GaAs substrate through the formation of misfit dislocations. We then deposit \InMnAs~(BEPR of As/In = 6, In/Mn = 8) at a low substrate temperature, $T_s=250$~$^\circ$C, to allow uniform incorporation of substitutional Mn and avoid the formation of MnAs clusters. The chosen In/Mn flux ratio is intended to yield an \InMnAs~composition in the range $x \sim 0.02 - 0.03$. Finally, we grow a Cd$_3$As$_2$ layer at $T_s= 100$~$^\circ$C. The growth temperatures are monitored using band edge thermometry. The surface is monitored {\it in situ} during growth using a 12 keV Staib NEK150R1 reflection high-energy electron diffraction (RHEED) system. 

\subsection*{STEM Sample Preparation and Experimental Details}

Cross-section samples for scanning transmission electron microscopy (STEM) analysis were prepared on an FEI Helios Nanolab G4 dual-beam focused ion beam (FIB) instrument. First, the sample was coated with amorphous carbon prior to cross-sectioning to protect the film surface from damage on ion-beam exposure. The sample was thinned using a $30$~keV Ga ion beam followed by a $2$~keV ion-beam shower to remove any surface damaged layers during thinning. High-angle annular dark-field (HAADF)-STEM imaging and STEM energy dispersive X-ray (EDX) mapping were performed on an aberration-corrected FEI Titan G2 $60-300$ (S)TEM microscope, which is equipped with a monochromator, CEOS-DCOR probe corrector and a super-X EDX spectrometer. The microscope was operated at $200$~keV with a probe convergence angle of $18.2$~mrad and HAADF detector collection angles of $55$ and $200$ mrad. EDX elemental maps and HAADF images were collected using a probe current of $100$~pA.

\subsection*{Electrical magnetoresistance measurements}
We carry out electrical magnetoresistance measurements on Hall bar devices with dimensions $100~\mu$m$\times$50~$\mu$m and $50~\mu$m$\times$25~$\mu$m. These are fabricated using standard photolithography (MLA150 Heidelberg) and Ar plasma etching. Ohmic contacts are formed using e-beam evaporation of $10/30$~nm Cr/Au. We fabricate a top-gate using a $30$~nm Al$_2$O$_3$ dielectric layer deposited using atomic layer deposition, followed by e-beam evaporation of a $10/30$~nm Cr/Au metallic contact. We perform the transport measurements in a Dynacool PPMS (Quantum Design) and Triton He-3 refrigerator (Oxford Instruments).

\section*{Data Availability}
The data associated with all the results presented in the main manuscript and in the Supplementary materials is available upon reasonable request to the corresponding author. It will also be accessible at Penn State's ScholarSphere repository (https://scholarsphere.psu.edu/about).

\section*{Code Availability}
All code for the calculations presented in the main manuscript and in the Supplementary information is described in Methods. Code packages used for first-principles computational work is available at https://www.vasp.at/. Any additional information is available upon reasonable request to the corresponding author.

\section*{Acknowledgments}

We acknowledge partial support for this project, including the use of facilities and instrumentation, from the Pennsylvania State University Materials Research Science and Engineering Center [National Science Foundation grant number DMR-2011839] (SI, ES, NS). ES acknowledges support from the National Science Foundation Graduate Research Fellowship Program under Grant No. DGE1255832. KJY and CXL acknowledge the support of the NSF grant award (DMR-2241327). BN and YW acknowledge startup funds from the University of North Texas, and computational resources from the Texas Advanced Computing Center. Part of the modeling was supported by computational resources from a user project at the Center for Nanophase Materials Sciences (CNMS), a US Department of Energy, Office of Science User Facility at Oak Ridge National Laboratory. This project was initiated in part using support from the Institute for Quantum Matter under DOE EFRC Grant No. DESC0019331 (JC, TM, NS). Parts of this work were carried out in the Characterization Facility, University of Minnesota, which receives partial support from the NSF through the MRSEC (DMR-2011401). SG acknowledges support from a Dissertation Fellowship from the graduate school at the University of Minnesota. SG and KAM were supported by and NSF Grant No. DMR-2309431.

\section*{Author contributions}

N.S. conceived of the project exploring the properties of \CdAs/\InMnAs~heterostructures. MBE growth and nanofabrication were carried out by E.S. and S.I. followed by transport measurements/analysis by S.I., all under the supervision of N.S. The theoretical calculations were carried out by K.Y. and B.N. under the supervision of C.L. and Y.W. The TEM measurements and analysis were carried out by S.G. under the supervision of K.A.M. The high purity \CdAs~source material was synthesized by J.C. and T.M.M. The writing of the manuscript was led by S.I. and N.S. with input from all authors.

\newpage

\newpage


\providecommand{\noopsort}[1]{}\providecommand{\singleletter}[1]{#1}%

\end{document}